\documentclass[twoside,11pt]{article}

\usepackage{jmlr2e}

\usepackage{amsmath}
\usepackage{enumerate}
\usepackage{url} 
\usepackage[utf8]{inputenc}
\usepackage{amsfonts}
\usepackage{float}
\usepackage{mathtools}
\usepackage{array}
\usepackage{multirow}
\usepackage{tikz}
\usepackage{bm}
\usepackage{booktabs}

\DeclareMathOperator*{\argmin}{argmin}

\usepackage{lastpage}

\ShortHeadings{Trees With Structured and Mixed-Type Covariates}{Giubilei, Padellini and Brutti}
\firstpageno{1}

\begin{document}

\title{Energy Trees: Regression and Classification With Structured and Mixed-Type Covariates}

\author{\name Riccardo Giubilei \email rgiubilei@luiss.it\\
        \addr Department of Statistics, Sapienza University of Rome, Rome, Italy\\
        \addr Department of Economics and Finance, Luiss Guido Carli, Rome, Italy\\
       \AND
       \name Tullia Padellini \email tullia.padellini@bancaditalia.it\\
        \addr DG Economics, Statistics and Research, Bank of Italy, Rome, Italy\\ 
        \AND
        Pierpaolo Brutti \email pierpaolo.brutti@uniroma1.it\\
        \addr Department of Statistics, Sapienza University of Rome, Rome, Italy}

\editor{}

\maketitle

\begin{abstract}
The increasing complexity of data requires methods and models that can effectively handle intricate structures, as simplifying them would result in loss of information. 
While several analytical tools have been developed to work with complex data objects in their original form, these tools are typically limited to single-type variables. In this work, we propose energy trees as a regression and classification model capable of accommodating structured covariates of various types. Energy trees leverage energy statistics to extend the capabilities of conditional inference trees, from which they inherit sound statistical foundations, interpretability, scale invariance, and freedom from distributional assumptions. We specifically focus on functional and graph-structured covariates, while also highlighting the model's flexibility in integrating other variable types. Extensive simulation studies demonstrate the model's competitive performance in terms of variable selection and robustness to overfitting. Finally, we assess the model's predictive ability through two empirical analyses involving human biological data. Energy trees are implemented in the R package \texttt{etree}.
\end{abstract}


\bigskip

\section{Introduction}\label{intro_section}

Increasingly often in data analysis, quantities of interest are complex objects living in non-Euclidean spaces, including curves, graphs, shapes, images, and strings. A popular approach to analyzing these objects is the translation into Euclidean feature vectors 
in order to apply ``standard'' statistical techniques \citep{jain2009}. The main limitation is that a universally valid way of obtaining such a representation does not exist \citep{jain2009}, 
leading to arbitrary choices and loss of information.


As the urgency of analyzing complex variables is growing, several frameworks for analyzing structured data without further 
simplification have been developed \citep{wang2007, jain2009, marron2014}. 
Particularly relevant is the case of object-oriented data analysis (OODA) \citep{wang2007}, which has been initially applied to tree-structured data objects \citep{wang2007}, becoming later successful in functional data analysis \citep{sangalli2009}. Additional examples in the field of structured data object analysis include graphs \citep{ginestet2017,zhou2022},
persistence diagrams \citep{bendich2016}, shapes \citep{dryden2016}, manifolds \citep{lila2016,lila2020}, sounds \citep{pigoli2018, tavakoli2019}, images \citep{benito2017}, covariance matrices \citep{dryden2009, pigoli2014}, and probability distributions \citep{chen2021,petersen2021}.

While ``analyzing data objects directly we avoid loss of information that occurs when data objects are transformed into numerical summary statistics'' \citep[p. 1]{larosa2016}, 
most existing contributions 
focus solely on single-type data objects. 
Consequently, the techniques are often domain-specific and cannot be easily extended to other data types. Even more critically, they do not support the joint analysis of multiple complex sources.

The pioneering work of \cite{balakrishnan2006} considers structured and mixed-type\footnote{In this article, \textit{mixed-type} does not refer to the dichotomy between numeric and categorical, but more generally to the case with \textit{covariates of different types}. As a further matter of terminology, numeric and categorical types are referred to as \textit{traditional}, which is intended as opposed to \textit{structured}, including any other type.} covariates, but it focuses on a single type of structured variables, requires domain-specific expertise, and lacks any concept of statistical significance.
\cite{brandi2018} and \cite{nespoli2019} also moved in the direction of a unifying learning framework for structured and mixed-type data, proposing two models that partially share building principles with this article. However, they have both proposed a very specific version that does not account for mixed-type data---not even including traditional types---and only allows for one type of structured covariates: functions in the first case, graphs in the second. Additionally, these works lack in-depth investigations into the model's structure, design principles, properties, and performance.

In this work, we introduce energy trees as a new and more general class of decision trees. The model has sound statistical foundations and provides a unifying framework to perform classification and regression with structured and mixed-type covariates. 
Energy trees draw essential 
features from conditional inference trees \citep{hothorn2006} and energy statistics \citep{szekely2013}. The tree structure facilitates the analysis of mixed-type variables, and the use of association tests, as in conditional trees, ensures statistically principled splits.
Energy statistics enable assessing the association between variables of different and possibly structured types. 

The article is organized as follows. Section \ref{energy_trees_section} describes the structure, the properties, and the algorithm of energy trees. Specifically, Section \ref{etree_varsel_section} and Section \ref{etree_split_section} focus on the two crucial steps of variable selection and splitting, respectively. Section \ref{etree_characterization_section} explains how to leverage the model's flexibility to accommodate any type of covariate. The properties and the performance of energy trees are demonstrated through various numeric applications with simulated (Section \ref{simulation_section}) and empirical (Section \ref{application_section}) data. Section \ref{conclusion_section} includes a brief recapitulation and ideas for future work.

\section{Energy Trees}\label{energy_trees_section}

Energy trees belong to the class of recursive partitioning models, or \textit{trees}. They share a similar structure with conditional trees \citep{hothorn2006}, which find splits employing permutation tests to estimate the conditional distribution of statistics measuring the association between dependent and explanatory variables. This approach allows overcoming two critical issues of traditional tree-based methods: selection bias and overfitting. 
In energy trees, the association is evaluated using energy tests of independence \citep{szekely2007} from the energy statistics framework. 
This fairly general class of tests allows assessing the association between variables defined in spaces that are not necessarily Euclidean and not necessarily the same.

Energy trees have several advantageous properties that derive from their constituent parts. As a tree-structured model, they are easily interpretable, scale-invariant, do not require preprocessing or distributional assumptions, can simply handle missing values, and provide automatic feature selection. Grounded in a conditional inference framework similar to that of conditional trees, they have statistically sound foundations, avoid selection bias, and exhibit robustness to overfitting. Finally, energy statistics enable the analysis of covariates that are potentially structured and of different types, positioning energy trees as a unifying framework for regression and classification with any kind of variable.

Energy trees are implemented in the R package \texttt{etree}, which is available on CRAN at \url{https://CRAN.R-project.org/package=etree} and on GitHub at \url{https://github.com/ricgbl/etree}.

\subsection{Structure and Algorithm}\label{etree_algorithm_section}

Energy trees take as input a learning sample $\mathcal{L}_n = \{ (Y_i, X_{1i}, \dots, X_{Ji}); \; i = 1, \dots, n \}$ consisting of a response variable with support $\mathcal{Y}$ and a set of covariates $\mathbf{X} = (X_1, \dots, X_J)$. For regression, $\mathcal{Y} \subseteq \mathbb{R}$; in the case of classification, 
$\mathcal{Y}$ is a discrete set of labels $\{1, \dots, K\}$. The domain of the $j$-th covariate, $j = 1, \dots, J$, is denoted with $\mathcal{X}_j$ and does not necessarily have to be a vector space. This means that the model can handle not only traditional variables but also structured ones in the form of complex objects such as strings, graphs, or functions. 

When growing any type of tree based on a generic $\mathcal{L}_n$, the observations $1, \dots, n$ are recursively partitioned into nodes that eventually determine which units have similar behavior regarding the response variable. 
Each node can be split into two or more kid nodes to whom observations are assigned based on the value taken in a single covariate. The initial node is called \textit{root}, while the ones without kid nodes are defined \textit{terminal nodes}. Each node in the tree is represented by a 
vector of \emph{case weights} $\mathbf{w} = (w_1, \dots, w_n)$, whose generic element $w_i$ is 
a non-negative integer that indicates the number of times the $i-th$ observation appears in the node.
Consequently, within node $\mathbf{w}$, the response variable is $Y^{\mathbf{w}} = ([Y_1]_{\times w_1}, \dots, [Y_n]_{\times w_n})$, and the $j$-th covariate is $X_j^{\mathbf{w}} = ([X_{j1}]_{\times w_1}, \dots, [X_{jn}]_{\times w_n})$, where$[a]_{\times b}$ indicates the repetition of element $a$ for $b$ times (or its absence if $b=0$). 

The recursive partitioning algorithm for energy trees can be summarized as follows:

\begin{enumerate}
\item \textbf{Stopping criterion.} For node $\mathbf{w}$, test the null hypothesis of global independence between the response variable $Y^{\mathbf{w}}$ and all the covariates $X_j^{\mathbf{w}} $, $j=1,\dots,J$, as $H_0=\cap_{j=1}^J H_0^j$, where the $J$ partial hypotheses $H_0^j: D(Y^{\mathbf{w}}|X_j^{\mathbf{w}})=D(Y^{\mathbf{w}})$ regarding the distribution $D(\cdot)$ are verified using energy tests of independence \citep{szekely2007}. If $H_0$ is not rejected at a pre-specified level $\alpha$, stop.
\item \textbf{Variable selection.} Select the $j^*$-th covariate $X_{j^*}^{\mathbf{w}}$ that exhibits the strongest association with the response $Y^{\mathbf{w}}$, based on the energy test of independence for $H_0^j$.
\item \textbf{Split.} Determine a non-empty set $A^* \subset \mathcal{X}_{j^*}$ to partition $\mathcal{X}_{j^*}$ into $A^*$ and $\mathcal{X}_{j^*} \setminus A^*$. The resulting case weights $\mathbf{w}_{\text{left}}$ and $\mathbf{w}_{\text{right}}$ define the two subgroups (one for each kid node) and have their elements computed as $w_{\text{left},i} = w_i \cdot I(X_{j^*i}  \in A^*)$ and $w_{\text{right},i} = w_i \cdot I(X_{j^*i} \notin A^*)$ for all $i = 1,\dots,n$, where $I(\cdot)$ is the indicator function. 
\item Repeat steps 1, 2 and 3 on 
nodes $\mathbf{w}_{\text{left}}$ and $\mathbf{w}_{\text{right}}$, respectively.
\end{enumerate}


\noindent Despite the similarity with the algorithm of conditional trees, two key differences enable the generalization to structured and mixed-type covariates. In step 1, the $J$ partial hypotheses $H_0^j: D(Y^{\mathbf{w}}|X_j^{\mathbf{w}})=D(Y^{\mathbf{w}})$ are verified using energy tests of independence instead of permutation tests; in step 3, determining the set $A^* \subset \mathcal{X}_{j^*}$ is remarkably more complicated.  Section \ref{etree_varsel_section} describes in greater detail steps 1 and 2, while Section \ref{etree_split_section} addresses step 3.

\subsection{Variable Selection}\label{etree_varsel_section}

Steps 1 and 2 of the algorithm involve verifying the $j$-th partial hypothesis $H_0^j: D(Y^{\mathbf{w}}|X_j^{\mathbf{w}})=D(Y^{\mathbf{w}})$ of independence between $Y^{\mathbf{w}}$ and $X_j^{\mathbf{w}}$
, for $j=1,\dots,J$. Each hypothesis is verified with an energy test of independence between the response and the $j$-th covariate, using the standard choice of $n\mathcal{V}_n^2$ as the test statistic, where $n$ is the sample size and $\mathcal{V}_n$ is the sample distance covariance \citep{szekely2007}. Thus, for each node $\mathbf{w}$, the sample distance covariance 
needs to be calculated between each covariate and the response.

Traditionally, the sample distance covariance is obtained by computing, for each variable, the Euclidean distance between couples of observations \citep{szekely2007, szekely2013}. However, energy trees allow for covariates in more complex spaces, requiring a generalized notion of distance between observations. Formally, consider node $\mathbf{w}$ and denote its size with $m$, i.e., $\sum_{i = 1}^n w_i = m$. The goal is to verify the $j$-th partial hypothesis $H_0^j: D(Y^{\mathbf{w}}|X_j^{\mathbf{w}})=D(Y^{\mathbf{w}})$.  Let $X_{jk}$, $k = 1, \dots, m$, be the generic element of $X_j^{\mathbf{w}}$, and $Y_k$, $k = 1, \dots, m$, that of $Y^{\mathbf{w}}$. Take a distance $\delta(\cdot)$ on the covariate's space $\mathcal{X}_j$ and the Euclidean norm $|\cdot|$ on $\mathcal{Y} \subseteq \mathbb{R}$. The observed test statistic $m\mathcal{V}_m^2(X_j^{\mathbf{w}}, Y^{\mathbf{w}})$ for node $\mathbf{w}$ is obtained as follows:

\begin{enumerate}
\item Compute the distance matrices defined by
\begin{equation*}
\begin{aligned}
(a_{kl}) &= \delta(X_{jk}, X_{jl}), \quad &k,l = 1, \dots, m, \\
(b_{kl}) &= |Y_k - Y_l|, \quad &k,l = 1, \dots, m;
\end{aligned}
\end{equation*}
\item For both matrices, calculate row means, column means, and global mean, e.g.,
\begin{equation*}
\bar{a}_{k\cdot} = \frac{1}{m} \sum_{l=1}^m a_{kl}, \qquad \bar{a}_{\cdot l} = \frac{1}{m} \sum_{k=1}^m a_{kl}, \qquad \bar{a}_{\cdot\cdot} = \frac{1}{m^2} \sum_{k,l=1}^m a_{kl};\\
\end{equation*}
\item Compute the centered distances, e.g.,
\begin{equation*}
\begin{aligned}
A_{kl} &= a_{kl} - \bar{a}_{k\cdot} - \bar{a}_{\cdot l} + \bar{a}_{\cdot\cdot}, \qquad &k,l = 1, \dots m; \\
\end{aligned}
\end{equation*}
\item Calculate the square of the sample distance covariance $\mathcal{V}_m(X_j^{\mathbf{w}}, Y^{\mathbf{w}})$ as
\begin{equation*}\label{etree_sample_dist_cov}
\mathcal{V}_m^2(X_j^{\mathbf{w}}, Y^{\mathbf{w}}) = \frac{1}{m^2} \sum_{k,l=1}^m A_{kl}B_{kl};
\end{equation*}
\item Compute the value of the observed test statistic for node $\mathbf{w}$ as $m\mathcal{V}_m^2(X_j^{\mathbf{w}}, Y^{\mathbf{w}})$.
\end{enumerate}

\noindent The procedure involves the calculation of $m(m-1)/2$ distances for both the covariate and the response variable, resulting in a computational time complexity of $O(m^2)$.

The sampling distribution of the test statistic $m\mathcal{V}_m^2(X_j^{\mathbf{w}}, Y^{\mathbf{w}})$ under the null hypothesis depends on the unknown joint distribution of $X_j^{\mathbf{w}}$ and $Y^{\mathbf{w}}$. Hence, it is estimated using permutation tests, that is, computing replicates of the test statistic under random reshuffles of the indices of $Y^\mathbf{w}$. The procedure is repeated for each covariate, yielding a p-value $P_j$ for each partial hypothesis $H_0^j: D(Y^{\mathbf{w}}|X_j^{\mathbf{w}})=D(Y^{\mathbf{w}})$. 

In step 1 of the energy trees algorithm, the test of global independence between the response variable $Y^{\mathbf{w}}$ and all the covariates $X_j^{\mathbf{w}}$, $j=1,\dots,J$, is formalized as $H_0=\cap_{j=1}^J H_0^j$. 
The p-value of the global test can be computed starting from $P_j$, $j = 1, \dots, J$, and using p-values adjustment techniques for multiple testing procedures. Energy trees adopt the false discovery rate (FDR) correction \citep{benjamini1995}. As opposed to Bonferroni correction---used in conditional trees---that controls the probability of at least one false rejection, FDR regulates the expected proportion of false rejections among all rejections, so it has less conservative control of type I error but also greater power. After correction, $H_0$ is rejected if the minimum of the adjusted p-values is less than a pre-specified nominal level $\alpha$; otherwise, the recursion stops. Hence, as in conditional trees, $\alpha$ may be interpreted not only as the nominal level controlling type I error in each node but also as a tunable hyperparameter determining the size of energy trees.

If the hypothesis of global independence is not rejected in step 1, the algorithm proceeds to step 2 (variable selection). The covariate selected for splitting is the one that yields the smallest p-value, i.e., $X_{j^*}^{\mathbf{w}}$ is such that $j^*=\text{argmin}_{j=1,\dots,J} P_j$.

\subsection{Splits}\label{etree_split_section}

The goal of step 3 is to use the splitting variable $X_{j^*}^{\mathbf{w}}$ to divide observations into two subgroups. Many methods can accomplish this when $X_{j^*}^{\mathbf{w}}$ is either numeric or categorical; Energy trees perform an energy test of independence for each possible split point. 
For numeric covariates, the independence test is conducted between the response variable and a binary vector that indicates which units would be assigned to the first kid node based on that split point. Note that the construction of the binary vector leverages the natural ordering among numeric values. In the case of nominal covariates, a natural ordering may not exist. Hence, the binary vector is obtained by considering each instance of all non-trivial combinations of the splitting variable's categories. In both cases, the optimal split point is the one yielding the strongest association in terms of p-value with the response. The idea of performing the test using a binary vector representing the split follows \cite{hothorn2006}, and it induces a two-sample statistic that maximizes the discrepancy between the distribution of the response in the two kid nodes.

Formally, the goal is to find the optimal $A^* \subset \mathcal{X}_{j^*}$, with $A^* \neq \emptyset$, to partition the splitting variable's domain $\mathcal{X}_{j^*}$ into two sets: $A^*$ and $\mathcal{X}_{j^*} \setminus A^*$. For traditional variables, $A^*$ is searched among the non-empty and proper subsets $Q$ of $\mathcal{X}_{j^*}$, where $Q$ belongs to a reasonable set $\mathcal{Q}$.
In energy trees, for each $Q$ in $\mathcal{Q}$, an energy test of independence is performed between the response variable and the binary vector given by 

\begin{equation}\label{split_binary}
\bm{\ell}_Q = \left( \big[ I \left( X_{j^*1} \in Q \right) \big]_{\times w_1}, \; \dots, \; \big[ I \left( X_{j^*n} \in Q \right) \big]_{\times w_n} \right).
\end{equation}

\noindent The optimal $A^*$ is selected as the $Q$ whose corresponding $\bm{\ell}_Q$ shows the strongest association in terms of p-value with $Y^{\mathbf{w}}$.

The form of $Q$ depends on the type of the splitting variable. In the numeric case, consider the $k$ unique values of $X_{j^*}^{\mathbf{w}}$ for which $w_i \neq 0$, and let $x_{(1)}, \dots, x_{(k)}$ be the corresponding sorted vector. The set $Q$ is a right-closed interval, $Q = (-\infty, q]$, where $q = x_{(1)}, \dots, x_{(k-1)}$. In other terms,

\begin{equation}\label{q_numeric}
\mathcal{Q} = \big\{ (-\infty, x_{(1)}], \dots, (-\infty, x_{(k-1)}] \big\}.
\end{equation}

In the nominal case, let $M = \{ 1, \dots, m \}$ represent the levels of $X_{j^*}^{\mathbf{w}}$ for which $w_i \neq 0$. In this case, $Q$ is any element of

\begin{equation}\label{q_nominal}
\mathcal{Q} = \mathcal{P}(M) \setminus \{ \emptyset, M \}, 
\end{equation}

\noindent where $\mathcal{P}(M)$ is the power set of $M$, and the trivial cases $Q = \emptyset$ and $Q = M$ are excluded. Moreover, due to the complementary nature of kid nodes' subspaces in binary partitioning models, additional cases can be ignored. 

The problem becomes more complicated if $X_{j^*}^{\mathbf{w}}$ is structured, such as in the case of curves, graphs, shapes, images, or strings. These types of observations do not have a natural ordering or an obvious way to split them into two subgroups. 
Therefore, alternative strategies for splitting must be considered, as discussed in the remaining part of this section.

\subsubsection{Feature vector extraction}\label{split_featvec_section}

The first presented method is called \textit{feature vector extraction}. It consists of finding the split after applying a transformation to switch from the complex sample space $\mathcal{X}_{j^*}$ of structured data objects to a more tractable Euclidean feature space. The name of the method derives from the popular practice to represent structured objects through Euclidean feature vectors \citep{jain2009}. 

In the specific context, the transformation depends on the type of the splitting variable and is a function $g_j: \mathcal{X}_{j^*} \rightarrow \mathbb{R}^{s_j}$. 
It involves expanding a structured covariate into real-valued coefficients, or components, similarly to \textit{basis expansion} in linear algebra or for the particular case of functional variables \citep{ramsay2005}. Hence, the transformation of the splitting variable into coefficients is referred to as \textit{coefficient expansion} in the following.

Let $X_{j^*k}$, $k = 1, \dots, m$, be the generic element of $X_{j^*}^{\mathbf{w}}$, and denote by $\bm{b}_{k}^j$ the $s_j \times 1$ vector resulting from the transformation $g_j$ applied to $X_{j^*k}$, i.e., $g_j(X_{j^*k}) \mapsto \bm{b}_{k}^j = (b_{k1}^j, \dots, b_{k s_j}^{j})^T$. If $X_{j^*}^{\mathbf{w}}$ consists of $m$ objects, the transposed collection of vectors $\mathbf{B}^j = [\bm{b}_{1}^j \, \cdots \, \bm{b}_{m}^j]^T$ is a $m \times s_j$ matrix. The matrix $\mathbf{B}^j$ can be also expressed as the collection of the components it contains, i.e., $\mathbf{B}^j = [\bm{b}_1^j \, \cdots \, \bm{b}_{s_j}^{j}]$; in other words, for $s = 1, \dots, s_j$, $\bm{b}_s^j = (b_{s1}^j, \dots, b_{sm}^j)^T$ is the $s$-th component resulting from the transformation of 
$X_{j^*}^{\mathbf{w}}$ via $g_j$. 

Feature vector extraction solves the issue of splitting structured covariates, but the resulting $\mathbf{B}^j=g_j(X_{j^*}^{\mathbf{w}})$ is an $m \times s_j$ matrix that includes the $s_j$ components $\bm{b}_s^j$, $s = 1, \dots, s_j$. To handle this multidimensional problem, the approach replicates the setup of multiple (originally) numeric covariates. First, the most associated component $\bm{b}_{s^*}^j$ is selected through an energy test of independence between the response variable $Y^{\mathbf{w}}$ and each component $\bm{b}_s^j$, $s = 1, \dots, s_j$. This process is as described in Section \ref{etree_varsel_section}, except that no stopping criterion is used here. Then, the split point for the selected real-valued $\bm{b}_{s^*}^j$ is determined similarly to numeric covariates. The optimal $A^* \subset \mathcal{X}_{j^*}$ is the subspace $Q$ corresponding to the binary vector $\bm{\ell}_Q$ from Equation \eqref{split_binary} that, among all $Q$ in $\mathcal{Q}$ as defined in Equation \eqref{q_numeric}, yields the strongest association with the response variable $Y^{\mathbf{w}}$ in an energy test of independence. 

While the use of feature vector extraction for splits may initially seem counterintuitive in a context where the goal is to analyze data in their original form, ``it can be quite hard to directly understand population structure using the object space alone. Thus, it is useful to simultaneously consider the (closely linked) feature space as well'' \citep[p. 734]{marron2014}. In other words, combining the two perspectives may lead to improved results and more meaningful interpretations \citep[ch. 3]{marron2021}.

\subsubsection{Clustering}\label{split_clustering_section}

Feature vector extraction is a valuable technique; however, it necessarily implies loss of information.
Since the splitting step implies partitioning the observations, a natural alternative strategy is 
to use clustering, as suggested by \cite{balakrishnan2006}. Specifically, distance-based clustering methods allow keeping the splitting variable in its original form and involve two steps: first, identify two (or more) representative data objects, or \textit{medoids}; second, assign other observations to the cluster that minimizes the distance from the corresponding medoid. 

Let $X_{j^*}^{\mathbf{w}}$ be the splitting covariate, $\mathcal{X}_{j^*}$ its sample space, and $\delta(\cdot)$ the distance defined on $\mathcal{X}_{j^*}$. The optimal $A^* \subset \mathcal{X}_{j^*}$, with $A^* \neq \emptyset$, for splitting $\mathcal{X}_{j^*}$ into two sets $A^*$ and $\mathcal{X}_{j^*} \setminus A^*$ is determined by first identifying two medoids $C_1$ and $C_2$, one for each kid node. Then, other observations are assigned to the kid node that corresponds to the closest medoid in terms of $\delta(\cdot)$. Specifically, the kid node to which the $k$-th element $X_{j^*k}$ of $X_{j^*}^{\mathbf{w}}$ is assigned is given by

\begin{equation}\label{centroids_det}
\argmin_{c \, \in \, \{C_1, C_2\}} \; \delta(X_{j^*k}, \, c).
\end{equation}

\noindent Note that Equation \eqref{centroids_det} 
implies that the optimal $A^*$ can be defined as the Voronoi region associated with one of the two medoids; e.g., for $C_1$,

\begin{equation}\label{voronoi}
A^* = \{ x \in \mathcal{X}_{j^*} : \delta(x, C_1) \leq \delta(x, C_2) \}.
\end{equation}

Many clustering techniques are based on finding representative observations among a set of structured data objects. Energy trees employ partitioning around medoids (PAM) \citep{kaufmann1987}, also known as \textit{k-medoids}. The computational complexity of PAM is $O(n^2)$, similarly to many other distance-based clustering algorithms. 
Among these, it is worth mentioning k-groups \citep{li2017}, which lies within the energy statistics framework. 
However, PAM is preferred because it has well-established faster variants such as CLARA \citep{kaufman2008} and FastPAM \citep{schubert2019}.

The clustering approach to splits offers the advantage of working directly with data objects without using arbitrary transformations. It enables the implementation of multiway splits, which may be particularly useful in multi-class problems, 
by considering more than two representative observations simultaneously. Its computational time is preferable to the $O(n^3)$ complexity of feature vector extraction. However, an important drawback is the lack of any concept of statistical significance for the splitting step. This is in contrast to feature vector extraction, where splits are based on statistical tests. Moreover, feature vector extraction provides enhanced interpretability by allowing focused analysis of specific aspects concerning single components.
The conclusion is that no dominant strategy exists, and the optimal approach should be determined application-wise. 

\section{Definition of Distances and Coefficient Expansion Methods}\label{etree_characterization_section}

Energy trees are designed to accommodate covariates of various types. In the following, \textit{traditional} refers to numeric and categorical variables, while \textit{structured} encompasses any other type, such as functional and in the form of graphs.
Each type of covariate requires an appropriate distance, which is necessary 
for computing the test statistic in the energy tests of independence employed for variable selection. It is also needed for structured covariates when using the clustering approach to splitting. Alternatively, when using feature vector extraction for splits, structured covariates necessitate a suitable method for coefficient expansion. Finally, any type of traditional covariate requires the distance for categorical variables to determine the split point, as outlined in Equation \eqref{split_binary}.

Table \ref{choices_what} provides the choices of distance and coefficient expansion for each type of covariate considered in this paper's applications. While standard methods apply for numeric, nominal, and functional variables, no natural technique exists in the literature when data are in the form of graphs \citep[p. 69]{marron2021}. Energy trees use the edge difference distance \citep{hammond2013}, defined as the Frobenius norm of the difference between the adjacency matrices of the two graphs, and the shell distribution \citep{carmi2007}, obtained via $k$-cores \citep{seidman1983}, $s$-cores \citep{eidsaa2013}, and $d$-cores \citep{giatsidis2013}, for binary, weighted, and directed graphs, respectively. Further details can be found in Appendix \ref{app:characterization}.

\begin{table}
\center
\begin{tabular}{p{0.15\textwidth}>
{\centering}p{0.22\textwidth}>
{\centering\arraybackslash}p{0.22\textwidth}}
\hline\noalign{\smallskip}
{} & \textbf{Distance} & \textbf{Coefficient exp.} \\
\noalign{\smallskip}\hline\noalign{\smallskip}
\textbf{Numeric} \quad & Euclidean & \textit{NA} \\
\textbf{Nominal} & Gower & \textit{NA} \\
\textbf{Functional} & $L^2$-norm & Cubic B-splines \\
\textbf{Graphs} & Edge difference & Shell distribution \\
\hline
\noalign{\smallskip}
\end{tabular}
\caption{Choices of distance and coefficient expansion for the four types of covariates considered in this work.}
\label{choices_what}
\vspace{-0.2cm}
\end{table}

One notable advantage of energy trees is their high flexibility. The choices of distance and coefficient expansion can be adjusted according to 
personal preferences or application-specific needs. Moreover, the model can easily accommodate other types of covariates as long as the corresponding distance is specified. This implies that many types of variables equipped with their own distance can be incorporated into the framework defined by energy trees. Examples include shapes and covariance matrices with Procrustes distance, manifolds with Riemannian distance, images with image Euclidean distance \citep{wang2005}, time series and sounds with dynamic time warping, probability distributions with $f$-divergences, strings with edit distances, and data objects in general metric spaces with Gromov-Haussdorf distance or 
fused Gromov-Wasserstein distance \citep{vayer2020}.
It is important to mention that distance covariance characterizes independence only for metric spaces of strong negative type \citep{lyons2013}, though it can be extended to semimetric spaces of negative type \citep{sejdinovic2013}. However, even when these conditions are not met, distance covariance can still be interpreted as a loose measure of association.

\section{Simulation Study}\label{simulation_section}

Energy trees are unbiased, robust to overfitting, and select meaningful covariates. This section presents three simulation scenarios that empirically validate these properties. The setup extends the original work on conditional trees \citep{hothorn2006} 
to the case of structured and mixed-type covariates. Experiments are conducted using $10,000$ replications and calculating $95\%$ confidence intervals through the normal approximation to the binomial distribution.

\subsection{Unbiasedness}\label{sim1}
In a recursive partitioning model, \textit{unbiasedness} refers to the selection of covariates $X_1, \dots, X_J$ with equal probabilities of $1/J$ under the null hypothesis of global independence between the response variable and predictors \citep{hothorn2006}.
To empirically demonstrate unbiasedness, it is necessary to verify if each covariate is selected with approximately the same relative frequency under independence. In this case, where no association exists between the response and the covariates, the root split is forced by removing any stopping criterion.
The response variable $Y$ follows a standard normal distribution $\mathcal{N}(0, 1)$, while the covariates are specified as:

\begin{enumerate}[\indent $X_1$.]
\item Numeric: uniformly distributed between $0$ and $1$;
\item Nominal: binary variable with uniformly-sampled values;
\item Functions: Gaussian random processes over $100$ evaluation points ranging from $0$ to $1$, with a mean of $0$ and the identity matrix as the covariance matrix;
\item Graphs: Erdős–Rényi random graphs with $100$ vertices and a connection probability of 0.2.
\end{enumerate}

\noindent Energy trees are compared to decision trees and conditional trees in terms of performance. However, since these models cannot handle structured covariates, feature vector extraction is employed to transform them into numeric components (see Section \ref{etree_characterization_section} and Appendix \ref{app:characterization} for details). Additional adjustments are explained in Appendix \ref{app:comparison}. The results of the scenario are presented in Table \ref{mixed_indep_table}. Decision trees are known to exhibit bias towards covariates with many possible splits: the least selected covariates are $X_2$, which takes only two values, and $X_3$, whose components have few split points. Conditional trees display far less bias but fail to include the reference probability $0.25$ within any of the approximated $95\%$ confidence intervals. Energy trees yield estimates that are very close to $0.25$ for all covariates and include this value within the approximated $95\%$ confidence intervals. They are unbiased regardless of the measurement scale or type of covariates. Consequently, energy trees stand out as the only model among the three that exhibits unbiasedness in the general case of structured and mixed-type covariates.

\vspace{0.4cm}

\begin{table}[H]
\centering
\begin{tabular}{lcccccc}
  \hline\\[-0.38cm]
    \multirow{2}[3]{5em}{\textbf{Covariate}\\[-0.2cm]} & \multicolumn{2}{c}{\textbf{Decision trees}} & \multicolumn{2}{c}{\textbf{Conditional trees}} & \multicolumn{2}{c}{\textbf{Energy trees}}\\[-0.06cm] \cmidrule(lr){2-3} \cmidrule(lr){4-5} \cmidrule(lr){6-7}\\[-0.5cm]
 & \textbf{Estimate} & \textbf{CI} & \textbf{Estimate} & \textbf{CI} & \textbf{Estimate} & \textbf{CI} \\[0.03cm]
  \hline\\[-0.4cm]
  $X_1$ (Numeric) & 0.4846 & \begin{tabular}{@{}c@{}}(0.4748, \,\\[-0.1cm] 0.4944)\end{tabular} & 0.2604 & \begin{tabular}{@{}c@{}}(0.2518, \,\\[-0.1cm] 0.2690)\end{tabular} & 0.2505 & \begin{tabular}{@{}c@{}}(0.2420, \,\\[-0.1cm] 0.2590)\end{tabular} \\
  $X_2$ (Nominal) & 0.0293 & \begin{tabular}{@{}c@{}}(0.0260, \,\\[-0.1cm] 0.0326)\end{tabular} & 0.2741 & \begin{tabular}{@{}c@{}}(0.2654, \,\\[-0.1cm] 0.2828)\end{tabular} & 0.2492 & \begin{tabular}{@{}c@{}}(0.2407, \,\\[-0.1cm] 0.2577)\end{tabular} \\
  $X_3$ (Functions) & 0.3954 & \begin{tabular}{@{}c@{}}(0.3858, \,\\[-0.1cm] 0.4050)\end{tabular} & 0.1901 & \begin{tabular}{@{}c@{}}(0.1824, \,\\[-0.1cm] 0.1978)\end{tabular} & 0.2503 & \begin{tabular}{@{}c@{}}(0.2418, \,\\[-0.1cm] 0.2588)\end{tabular} \\
  $X_4$ (Graphs) & 0.0907 & \begin{tabular}{@{}c@{}}(0.0851, \,\\[-0.1cm] 0.0936)\end{tabular} & 0.2754 & \begin{tabular}{@{}c@{}}(0.2666, \,\\[-0.1cm] 0.2842)\end{tabular} & 0.2500 & \begin{tabular}{@{}c@{}}(0.2415, \,\\[-0.1cm] 0.2585)\end{tabular} \\ 
\hline
\end{tabular}
\caption{Simulated point estimates and approximated $95\%$ confidence intervals for the relative frequencies of variable selection under independence between the response and the covariates when no stopping criterion is applied.
}
\label{mixed_indep_table}
\end{table}

\subsection{Overfitting and Selection of Meaningful Covariates}\label{sim2}
In recursive partitioning models that employ statistical tests of independence for variable selection, \textit{power} represents the probability of selecting any covariate, rather than stopping, under the alternative hypothesis of association with the response. Power can be examined by analyzing the behavior of the probability of selecting any variable (without forcing the split) as the association between the response and a specific covariate increases. When independence holds, the probability should be close to zero. 
As the association grows, it can be properly interpreted as the power of the independence test, so larger values indicate higher power.
Another relevant quantity is the \textit{conditional probability} of selecting the associated covariate, given that any variable is chosen for splitting. In the case of independence, each covariate should have an equal probability of being selected, resulting in a conditional probability of $1/J$ for all $j = 1, \dots, J$. As the association between the response and a covariate grows, the conditional probability should also increase. The power analysis assesses the robustness to overfitting, focusing on the ability to perform splits only when necessary. On the other hand, the conditional probability analysis concentrates on the selection of meaningful covariates. 

\enlargethispage{\baselineskip}

The simulation scheme for both analyses is similar to Section \ref{sim1}, except for $Y$ and one covariate. The response variable $Y$ follows a normal distribution with unit variance and a mean $\mu = 0$ for half the observations, while the other half have $\mu \in [0,1]$. The association between the response and one explanatory variable is induced by increasing the value of $\mu$ within the interval $[0,1]$. Since the focus of this work is on structured covariates, the associated predictor is initially the functional variable $X_3$ and then the graph-structured variable $X_4$. In the first case, $X_3$ is defined using the same two groups of $Y$: half of the observations are realizations of a Gaussian random process over $100$ evaluation points from $0$ to $1$, with a mean of $0$ and the identity matrix as the covariance matrix, while the other half has a mean of $0.5$. In the second case, 
half of the observations are Erdős–Rényi random graphs with $100$ vertices and a connection probability of $0.2$, while the other half have a connection probability of $0.8$. When the associated covariate is $X_j$, where $j = 3, 4$, the configuration of $X_i$, for $i =  1, \dots, 4$ and $i \neq j$, is the same as in Section \ref{sim1}.

\begin{figure}
\center
\hspace*{-1.3cm}
\includegraphics[width = 1.17\textwidth]{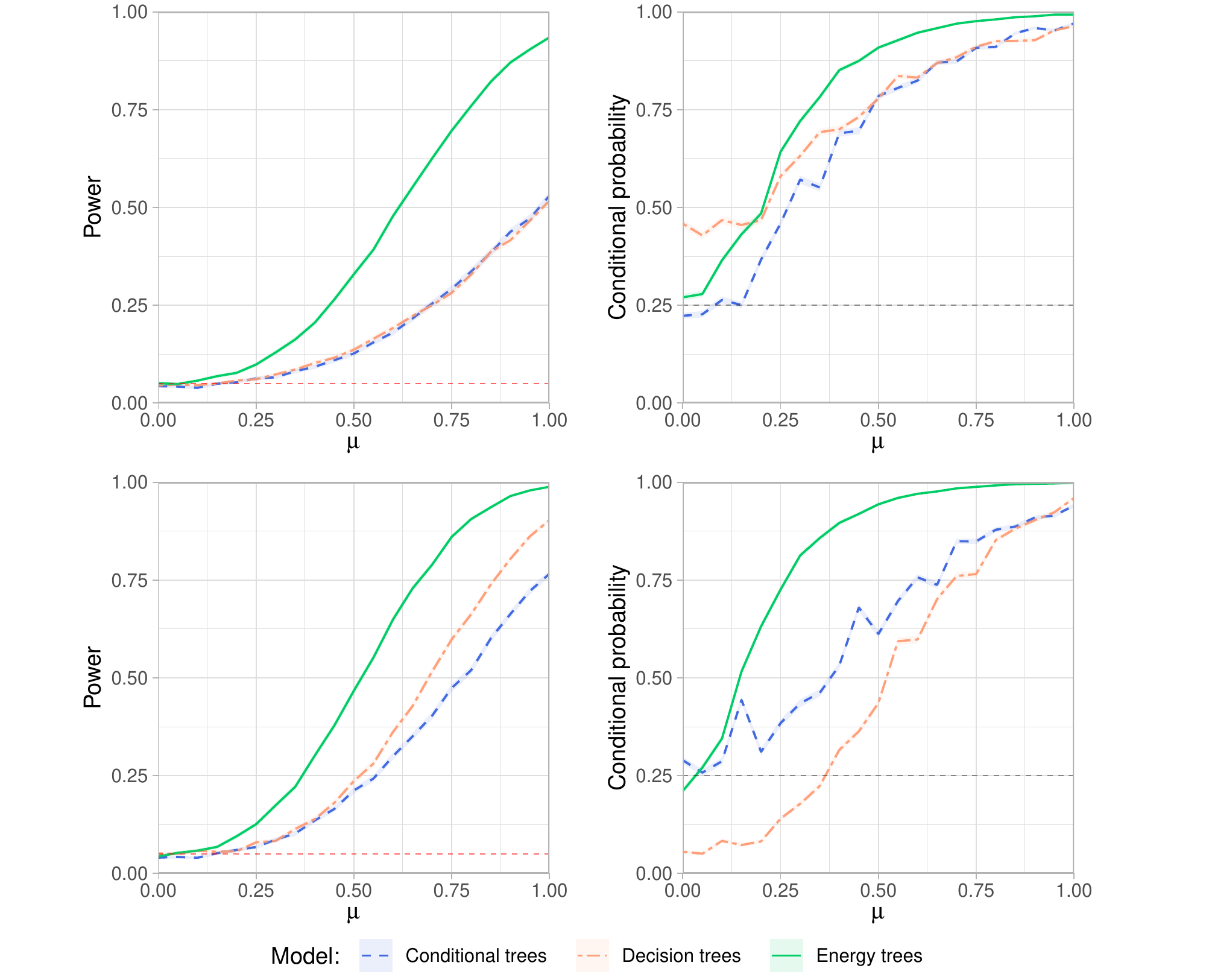}
\caption{Simulated power and conditional probability curves using four covariates of different types. First row: functional covariate associated with the response. Second row: graph-structured covariate associated with the response. For each model, point estimates and approximated $95\%$ confidence intervals are shown. The horizontal dashed lines represent the significance level of $\alpha = 0.05$ on the left and the reference probability of $0.25$ on the right.
}
\label{assfungph_plot}
\end{figure}

The results of the power and conditional probability analyses for decision trees, conditional trees, and energy trees are presented in Figure \ref{assfungph_plot}.
When the associated covariate is $X_3$, the estimated probability of selecting any covariate for $\mu = 0$ is bounded above by the critical threshold $\alpha = 0.05$ 
for conditional trees ($0.0432$) and decision trees ($0.0453$), and approximately for energy trees ($0.0507$). However, the power curve for energy trees widely dominates the other two models across the entire range.
The conditional probability for $\mu = 0$ closely approximates the reference value of $0.25$ only for energy trees ($0.2702$), while it is lower for conditional trees ($0.2231$) and substantially higher for decision trees ($0.4585$). In this case, the curve for energy trees is uniformly greater than the others since $\mu = 0.2$; before this value, the superiority of decision trees is artificially induced by their bias. When the associated covariate is $X_4$, the estimated probability of selecting any covariate for $\mu = 0$ is bounded above by the critical threshold $\alpha = 0.05$ 
only for conditional trees ($0.0406$) and energy trees ($0.0436$), and approximately for decision trees ($0.0511$). As the association increases, the power curve for energy trees dominates the other two. The conditional probability for $\mu = 0$ is relatively close to the reference value of $0.25$ for energy trees ($0.2110$) and conditional trees ($0.2894$), while it is substantially lower for decision trees ($0.0554$). When the response is associated with the graph-structured covariate, the conditional probability curve for energy trees is uniformly higher than the other two models.

For both types of structured covariates, energy trees limit the proportion of incorrect decisions in the root node to $\alpha$ when the response is independent of the predictors. Additionally, in this case, the proportion of selection for the covariate of interest given an incorrect split is approximately equal to the reference probability of $0.25$. When the response is associated with one of the covariates, energy trees exhibit higher power and more frequently select the correct covariate (excluding low levels of association) compared to the competitors. Consequently, the two analyses demonstrate that energy trees are not only robust to overfitting and capable of selecting meaningful variables when dealing with structured and mixed-type covariates but also preferable to the other two models in these respects.

\section{Real-World Data Applications}\label{application_section}

The predictive ability of energy trees is validated through two empirical analyses using real-world data. The first one focuses on a classification task that aims to detect knee osteoarthritis based on bones' shape and demographic information. The second analysis is a regression problem whose goal is predicting the intelligence quotient using multimodal brain connectomes and demographic information. These analyses not only substantiate the flexibility of the energy tree model but also demonstrate its wide applicability across various fields, including human biology and medicine.

\subsection{Knee Osteoarthritis Classification}
\label{bones_section}

\begin{figure}
    \centering
    \includegraphics[width = 0.95\textwidth]{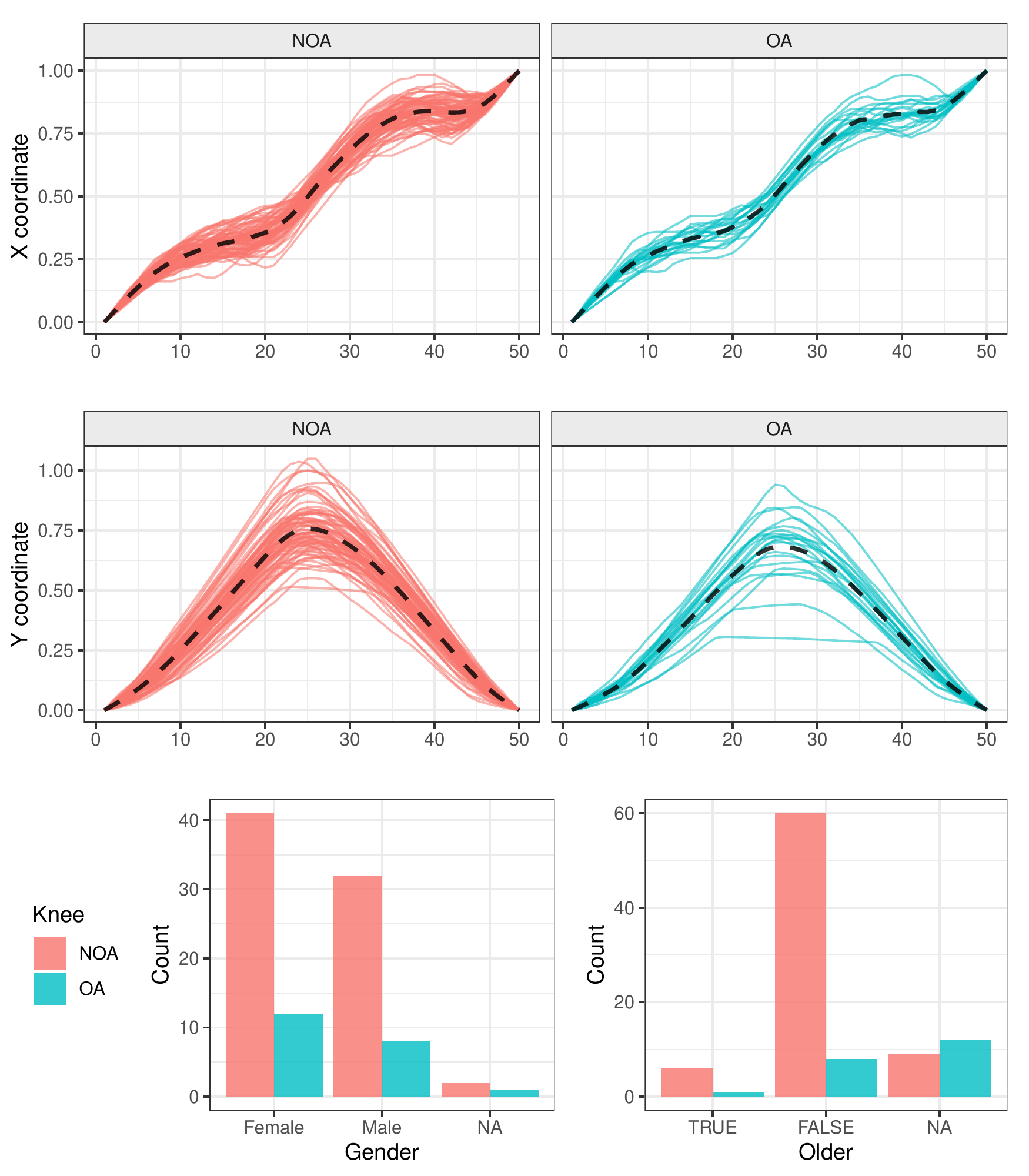}
    \vspace{0.2cm}
    \caption{Covariates used for knee OA classification. First row: observed values of the functional variable for the X coordinate (solid lines) and mean curves for the two groups (dashed lines). Second row: same, but for the Y coordinate. Third row: bar plots for nominal variables gender and older.}
    \label{bones_data_figure}
\end{figure}

Knee osteoarthritis (OA) is a painful and debilitating condition with a poorly understood etiology. However, the shape of the bones is a relevant risk factor because it directly influences the biomechanics of the joint.
Two previous studies \citep{shepstone2001, ramsay2007} used the shape of the femur's intercondylar notch, a deep fossa between two protrusions on the femur end closer to the knee joint, to distinguish between $21$ OA femora and $75$ non-OA (NOA) control femora. In the binary classification task conducted by \cite{ramsay2007}, the shape of the $j$-th intercondylar notch was transformed into two functional covariates, $X_j(t)$ and $Y_j(t)$, representing the longitudinal and latitudinal coordinates. These covariates were discretized to capture functional values 
at $50$ equally-spaced points ($t_1, \dots, t_{50}$) along the curves. The data set (see Figure \ref{bones_data_figure}) also includes the age and gender for each of the $96$ notches. Age is expressed as a binary variable, indicating whether the individual is older than $45$ or not, and gender is also binary.

\cite{ramsay2007} considered two models that can only handle functional data, hence they discarded the two nominal covariates. The first model, functional linear discriminant analysis (FLDA), treated the $100$ coordinates along the two curves as numeric variables, performed principal component analysis to reduce feature dimensionality, and used linear discriminant analysis on the resulting components to classify knees. The second model, mean difference projection (MPD), calculated a mean curve for each coordinate and each subpopulation (OA and NOA), and projected all the data onto the direction of the difference between the mean curves. Energy trees are implemented using feature vector extraction as the splitting method, 
and considering different combinations of significance level $\alpha$ and minimum number $\nu$ of observations in each terminal node: $\alpha \in \{ 0.1, 0.2, \dots, 1 \}$ and $\nu \in \{ 1, 5, 10 \}$. To ensure comparability, energy trees are tuned and evaluated using leave-one-out cross-validation (LOO-CV), similarly to \cite{ramsay2007}. Initially, only the two functional covariates are used for fitting. The results, in terms of binary classification performance metrics with OA cases as the positive class, are presented in Table \ref{table_bones_onlyfun}. Energy trees with $\alpha = 0.9$ and $\nu = 10$ outperform competitors in terms of accuracy, specificity, and positive predicted value (PPV). However, they perform worse than FLDA for sensitivity, negative predicted value (NPV), and balanced accuracy. In other words, when using only the two functional covariates, energy trees are the best model for correctly classifying NOA knees (at the expense of identifying only $43\%$ of OA knees and making $20\%$ overall errors), while FLDA is the best model for recognizing OA knees (at the expense of incorrectly classifying as OA more knees than those correctly identified and making $27\%$ overall errors).

While FLDA and MPD can only handle functional covariates and were specifically designed for the particular task, one of the greatest advantages of energy trees is that they provide a unifying framework for supervised learning with structured and mixed-type data. Considering the two nominal covariates as well, energy trees with both splitting methods outperform the two competitors for all the metrics (see Table \ref{table_bones_mixed}). The best parameter combination is $\alpha = 0.5$ and $\nu = 5$ for feature vector extraction, and $\alpha = 0.9$ and $\nu = 10$ for clustering. In the specific case, the two splitting methods focus on different aspects: feature vector extraction is preferable for recognizing OA cases, while clustering correctly classifies a larger number of NOA cases. 
Another study \citep{balakrishnan2006}, which employed decision trees allowing for functional covariates by performing clustering-based splits, analyzed the full set of covariates\footnote{Cf. \url{http://archive.dimacs.rutgers.edu/Research/MMS/PAPERS/fdt17.pdf}.}. Their model achieved an accuracy of $80.21\%$, which is comparatively worse than the $82.29\%$ and $84.38\%$ achieved by energy trees using the two splitting strategies.

\vspace{0.6cm}

\begin{table}[H]
\centering
\begin{tabular}{lcccccc}
\hline
& \textbf{Acc.} & \textbf{Sens.} & \textbf{Spec.} & \textbf{PPV} & \textbf{NPV} & \textbf{B.Acc.}\\
\hline
FLDA & $72.92$ & $66.67$ & $74.67$ & $42.42$ & $88.89$ & $70.67$ \\
MPD & $64.58$ & $57.14$ & $66.67$ & $32.43$ & $84.75$ & $61.90$ \\
ET & $80.21$ & $42.86$ & $90.67$ & $56.25$ & $85.00$ & $66.76$ \\
\hline
\end{tabular}
\caption{Performance metrics ($\%$) in LOO-CV for FLDA, MPD, and energy trees (ET) using only the two functional covariates.}\label{table_bones_onlyfun}
\end{table}


\vspace{-0.3cm}

\begin{table}[H]
\centering
\begin{tabular}{lcccccc}
\hline
& \textbf{Acc.} & \textbf{Sens.} & \textbf{Spec.} & \textbf{PPV} & \textbf{NPV} & \textbf{B.Acc.}\\
\hline
ET (FVE) & $82.29$ & $66.67$ & $86.67$ & $58.33$ & $90.28$ & $76.67$ \\
ET (C) & $84.38$ & $57.14$ & $92.00$ & $66.67$ & $88.46$ & $74.57$\\
\hline
\end{tabular}
\caption{Performance metrics ($\%$) in LOO-CV for ET with feature vector extraction (FVE) and clustering (C) as the splitting strategies, using the full set of covariates.}\label{table_bones_mixed}
\end{table}

\begin{figure}[h]
\vspace{-0.4cm}
    \center
    \includegraphics[width = \textwidth]{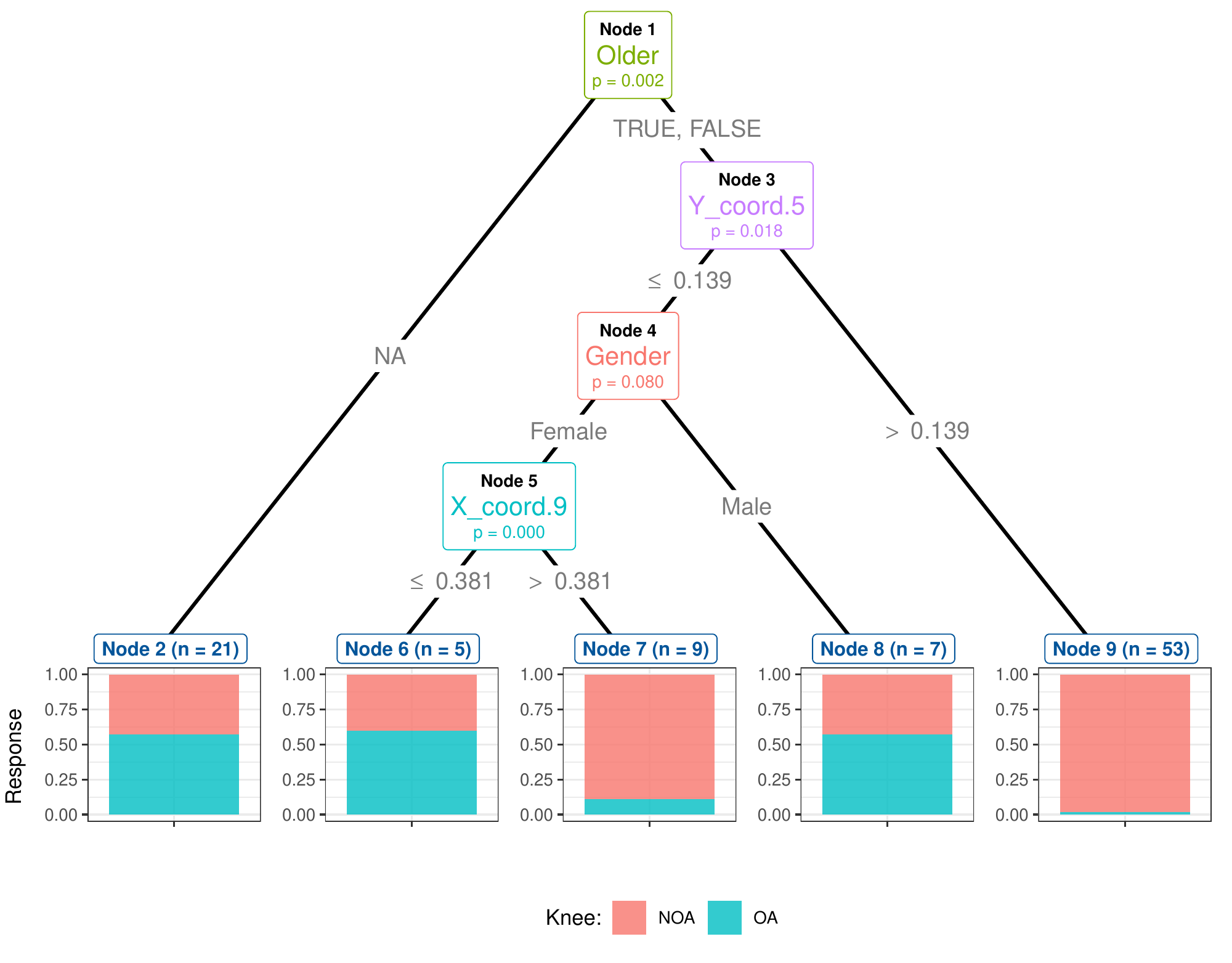}
    \vspace*{-0.5cm}
    \caption{Example of classification energy tree fitted in LOO-CV using feature vector extraction as the splitting method.}
    \label{bones_fitted_figure}
\end{figure}

An example of a single energy tree fitted through LOO-CV using the full set of covariates and feature vector extraction as the splitting method is given in Figure \ref{bones_fitted_figure}. The component selected for the split in the functional covariate representing the Y coordinate is the $5$-th, while the split in the variable denoting the X coordinate uses the $9$-th. These are the only two splits directly producing terminal nodes with a different balance between OA and NOA cases.

\subsection{Connectome-IQ Regression}
\label{nki_section}

\begin{figure}
    \centering
    \includegraphics[width = 0.85\textwidth]{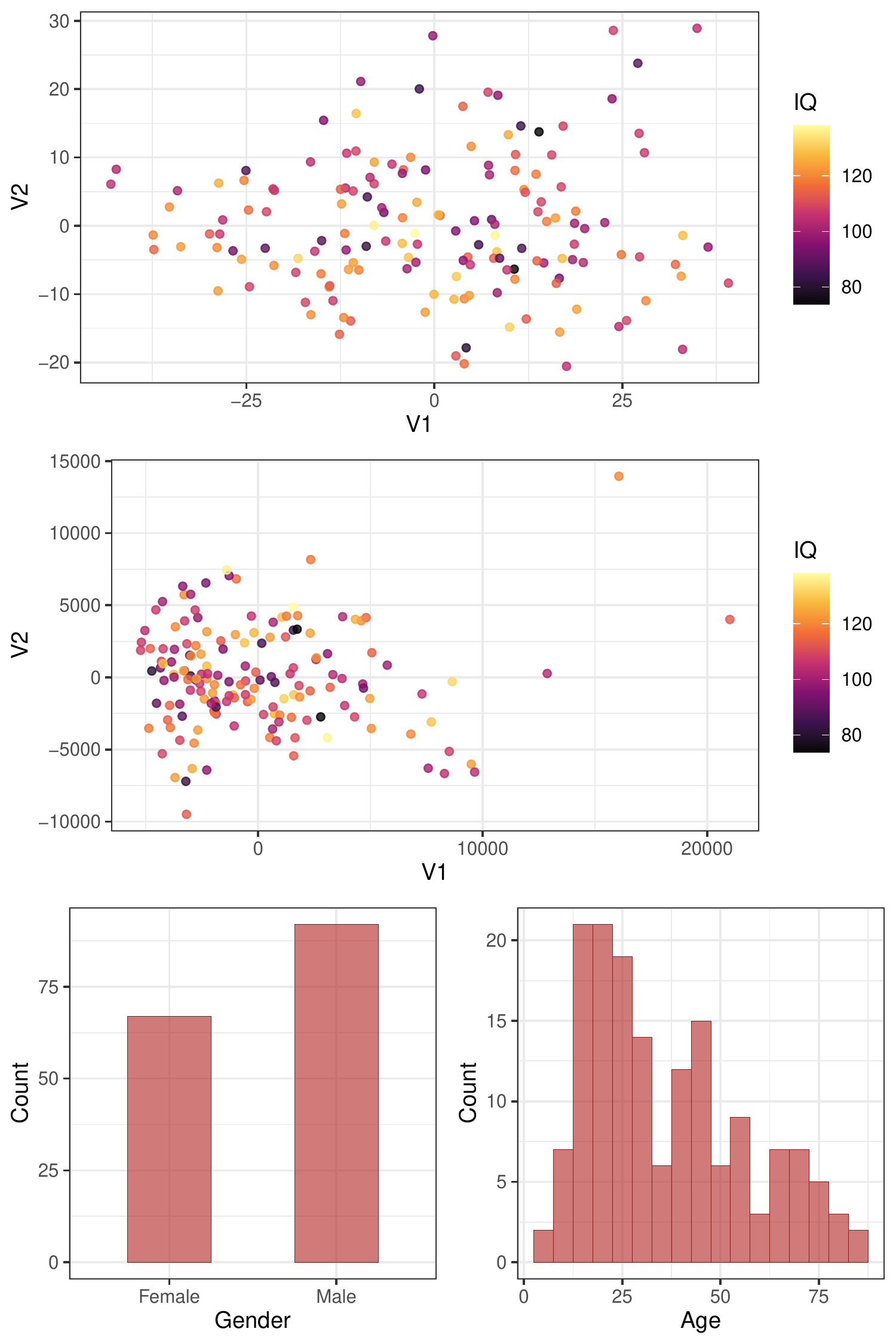}
    \caption{Covariates used in the regression analysis. First two rows: multidimensional scaling plots for the functional and structural graph covariates, respectively. Third row: bar plot for the nominal variable gender and histogram for the numeric variable age.}
    \label{nki_data_figure}
\end{figure}

Intelligence 
is commonly measured using tests that provide an intelligence quotient (IQ) score. 
Extensive research has demonstrated an association between high IQ scores and the coordinated activation of multiple brain regions, as observed through both structural \citep{haier2004,jung2007} and functional \citep{gray2003,lee2006} neuroimaging techniques.
Recent 
studies have investigated the neural basis of intelligence explicitly at the connectivity level \citep{li2009,hilger2017,dubois2018}. Inspired by these findings, the current analysis aims to investigate the regression relationship between IQ scores and various variables including structural and functional connectomes. The data used in this study derive from the Rockland-sample study \citep{nooner2012} conducted by the Nathan Kline Institute. Collecting and matching information from various sources, we have formed a data set with $159$ observations and four covariates (see Figure \ref{nki_data_figure}). Two of the covariates are graph-structured, representing the functional and structural connectomes of individuals as undirected and weighted graphs with $188$ nodes each. The remaining two covariates are traditional variables: age (with a mean of $36.43 \pm 20.09$) and gender ($67$ females and $92$ males). The response variable is the full-scale IQ, a scalar measure obtained for each subject in the Wechsler abbreviated scale of intelligence test, with a mean of $109.74 \pm 12.97$.

Similarly to Section \ref{simulation_section}, energy trees are fitted and compared 
with two other recursive partitioning models: decision trees and conditional trees. These competitors cannot handle structured covariates, hence requiring the transformation of graphs into feature vectors, which is performed using the shell distribution based on $s$-cores (see Section \ref{etree_characterization_section} and Appendix \ref{app:characterization} for details). 
Energy trees are implemented using both splitting methods. 
For energy trees and conditional trees, the stopping criteria $\nu$ and $\alpha$ are tuned, while decision trees use $\nu$ and a complexity parameter $cp$ that plays a role similar to $\alpha$ in determining the tree size.
To evaluate the models in an unbiased way, a nested $5$-fold cross-validation (CV) procedure is adopted: the inner folds are used for parameter tuning, and the outer folds for performance assessment.
The optimal parameter combination is selected based on the root mean square error (RMSE), averaged across the inner folds. The parameter values explored for the three models are the following: $\nu \in \{ 4, 7, \dots, 25 \}$, $\alpha \in \{ 0.05, 0.08, \dots, 0.20 \}$, and $cp \in \{ 0.005, 0.008, \dots, 0.020 \}$.
Table \ref{nki_cv} presents the results in terms of the RMSE over the outer folds.
The performance of energy trees with the two splitting methods is identical, and it results in an improvement of $2\%$ over decision trees and of $0.5\%$ over conditional trees.

\vspace{0.5cm}

\begin{table}[H]
\centering
\begin{tabular}{lcccccc}
\hline
 & Fold 1 & Fold 2 & Fold 3 & Fold 4 & Fold 5 & Average \\ 
\hline
ET (FVE) & $11.69$ & $13.47$ & $13.98$ & $11.57$ & $13.70$ & $12.88$ \\
ET (C) & $11.69$ & $13.47$ & $13.98$ & $11.57$ & $13.70$ & $12.88$ \\ 
DT & $13.58$ & $13.28$ & $13.52$ & $11.57$ & $13.91$ & $13.14$ \\ 
CT & $11.69$ & $13.47$ & $13.98$ & $11.87$ & $13.70$ & $12.94$ \\ 
\hline
\end{tabular}
\caption{RMSE over the outer folds of the nested $5$-fold CV for energy trees with both splitting methods, decision trees (DT), and conditional trees (CT).}
\label{nki_cv}
\end{table}

\begin{figure}[h]
    \center
    \includegraphics[width = \textwidth]{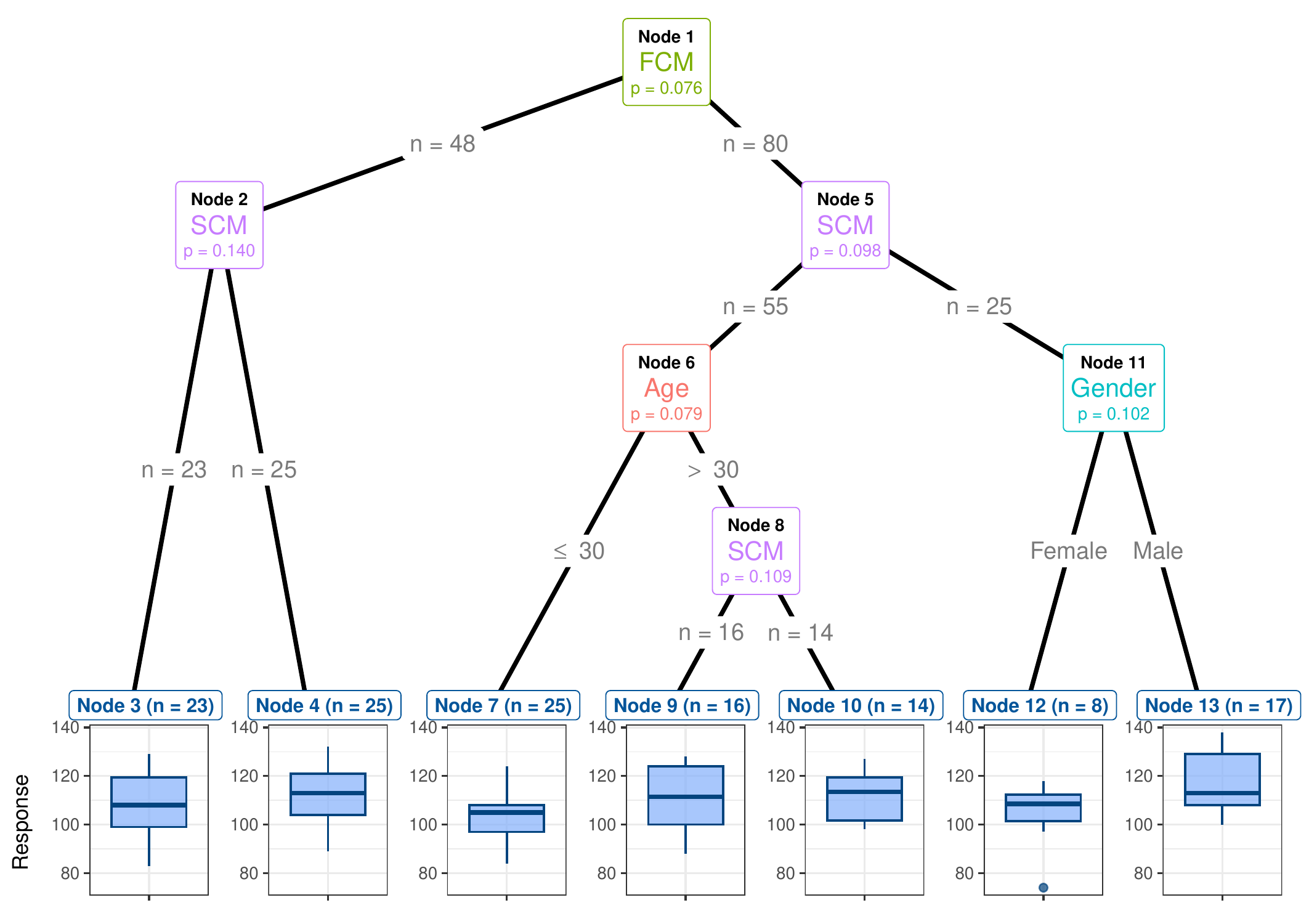}
    \vspace*{-0.7cm}
    \caption{Example of regression energy tree fitted using the outer folds of the nested 5-fold CV and using clustering as the splitting method.}
    \label{nki_fitted_figure}
\end{figure}

Figure \ref{nki_fitted_figure} illustrates an example of a single energy tree fitted using the outer folds of the nested 5-fold CV, with clustering as the splitting method. The tree hierarchy and the relative frequency of splits concerning graph-structured covariates confirm the relevant role of brain connectomes in the predictive task under consideration.

\section{Concluding Remarks}\label{conclusion_section}

Energy trees offer a unifying framework for classification and regression with structured and mixed-type data. They possess several advantageous properties derived from their constituent elements. As a recursive partitioning model, they are interpretable, scale-invariant, do not require preprocessing or parametric assumptions, have built-in feature selection, and can handle missing values, as well as covariates of different types. Using independence tests for variable selection and possibly for splitting ensures statistically sound foundations. The incorporation of energy statistics enables analyzing structured covariates. Furthermore, the simulation study has shown that energy trees exhibit unbiasedness, robustness to overfitting, and the ability to identify meaningful covariates for splitting. The model's predictive ability has been validated through two different experimental settings using real-world data. 
The ultimate advantage of energy trees lies in their great flexibility. This enables users to 1) choose between two alternative splitting strategies, 2) change the distance or the coefficient expansion technique as required, and 3) 
accommodate any other type of covariate. 

The simulation and empirical analyses not only confirm the need, as mentioned in Section \ref{intro_section}, for a model that can handle multiple complex sources, but also demonstrate that energy trees have competitive predictive performance compared to traditional models. Two main reasons account for this outcome: firstly, the ability to leverage all available data and retain each covariate, as shown in Section \ref{bones_section}, where it has made a significant impact; secondly, the decision to analyze covariates in their most natural form, avoiding the loss of information that occurs when transforming them into feature vectors, which has instead affected competitors in Section \ref{simulation_section} and Section \ref{nki_section}.
The simulation scenarios have proven that energy trees are unbiased regardless of the measurement scale or the type of covariate, unlike decision trees (which are biased) and conditional trees (which are only unbiased when working with traditional types). Additionally, energy trees display greater power and more frequently select the correct covariate across various levels of association compared to the other two models. The two empirical analyses are just two examples of the wide applicability of energy trees. The model could be also employed for image recognition, sentiment analysis, disease diagnosis, fraud detection, market segmentation, credit scoring, recommendation systems, and event detection, as well as for regression tasks in econometrics, finance, environmental sciences, geostatistics, social sciences, biostatistics, climate sciences, sports analytics, psychology, genetics, artificial intelligence, and many other fields.
The two splitting approaches, although structurally different, have demonstrated equal validity and may even yield identical results (see Section \ref{nki_section}). Yet, they may focus on different aspects (see Section \ref{bones_section}), so it is preferable to compare the two and select the most suitable one based on the specific application's characteristics and goals. 
Finally, the parameter $\alpha$ can be determined in a data-dependent way, as shown in Section \ref{application_section}, if prediction accuracy is the primary focus. Nevertheless, Section \ref{simulation_section} has proven that the classical value of $\alpha = 0.05$ performs well compared to potential competitors.

Much room is left for improvement. The theoretical derivation of the statistical properties of energy trees is desirable but challenging because the setting naturally involves diverse complex objects and mathematical spaces. Conducting a sensitivity analysis for input parameters would be crucial to gain a deeper understanding of the model's internal mechanisms and potentiality. The case of a structured response variable would be 
straightforward to implement in many respects, such as replacing the Euclidean norm with an appropriate distance, but developing a meaningful approach for summarizing each type of structured variable is essential to enable predictions. Conversely, the potential for incorporating new types of structured covariates and exploring novel applications is virtually limitless. Finally, it would be worthwhile to investigate and analyze ensemble methods, such as boosting, bagging, and random forests, that use energy trees as base learners.

\section*{Data Availability}

The data supporting the findings of the analysis in Section \ref{bones_section} can be openly accessed on the companion website to \cite{ramsay2007} at \url{http://www.stats.ox.ac.uk/~silverma/fdacasebook/notchchap.html}. IQ scores and phenotypical information of participants for the analysis in Section \ref{nki_section} are available at \url{http://fcon_1000.projects.nitrc.org/indi/pro/nki.html}, which is the Rockland-sample study page on the 1000 Functional Connectomes Project site managed by NeuroImaging Tools and Resources Collaboratory (NITRC). Raw DTI and fMRI neuroimaging data can be downloaded at \url{https://www.nitrc.org/frs/?group_id=404}, which is the File Release Download page on the NITRC site, under the section labeled \textit{Nathan Kline Institute}. The preprocessed version of this data---in the form of structural and (no GSR) functional connectomes \citep{brown2012}, as used in this paper---was previously available online and can now be shared upon reasonable request with the permission of the original authors \citep{brown2012}.

\section*{Disclosure Statement}

The authors declare no funding support or competing interests to disclose. The views expressed in this paper solely belong to the authors and do not involve any responsibility of the Bank of Italy and/or the Eurosystem.

\appendix

\section{Distance and Coefficient Expansion for Selected Types of Covariates}
\label{app:characterization}

Section \ref{etree_characterization_section} and Table \ref{choices_what} present the distances and coefficient expansion methods for the four types of covariates considered in this paper: numeric, nominal, functional, and graph-structured. Further details and mathematical formulations are provided below.

\subsection{Distance}

Each type of covariate requires a specific distance to compute the test statistic in the energy tests of independence used for variable selection. Additionally, traditional covariates need the distance for categorical variables to search for the split point, while structured covariates necessitate their own distance when performing splits with the clustering approach.

For traditional types numeric and nominal, energy trees use Euclidean and Gower's distance, respectively. Let $X_j$ be a numeric variable, meaning that observed values $X_{jk}$ and $X_{jl}$ are scalars. The distance $\delta_1(\cdot)$ between numbers is the Euclidean distance

\begin{equation*}
    \delta_1(X_{jk}, X_{jl}) = |X_{jk} - X_{jl}|.
\end{equation*}

If $X_j$ is a nominal variable, $X_{jk}$ and $X_{jl}$ are categories from a discrete set $\{ 1, \dots, K \}$. The distance $\delta_2(\cdot)$ between nominal data objects is the Gower's distance

\begin{equation*}
    \delta_2(X_{jk}, X_{jl}) = I(X_{jk} \neq X_{jl}).
\end{equation*}

When $X_j$ is functional, the $i$-th observed value $X_{ji}$ is a well-defined curve $f_i(t)$ where $t \in \mathcal{T}$ represents the evaluation points. The distance $\delta_3(\cdot)$ between functional observations $X_{ji} = f_i(\cdot)$ and $X_{jl} = f_l(\cdot)$ is the $L^2$ norm

\begin{equation*}
    \delta_3 (f_i, f_l) = || f_i - f_l ||_2 = \left( \int_{\mathcal{X}} | f_i(t) - f_l(t) | ^ 2 dt \right) ^ {1 / 2},
\end{equation*}

\noindent where $\mathcal{X}$ is the space over which $f_i(\cdot)$ and $f_l(\cdot)$ are defined. 

If $X_j$ is a graph-structured variable, the $i$-th observed value is a graph $G_i$. Any graph $G$ can be denoted as $G = (V, E)$, where $V$ is a set of vertices and $E$ is a set of edges. Each pair of vertices $u, v \in V$ can be represented as $e = \{ u, v \}$ and has an associated edge weight $w_e$, which is non-zero if and only if $e$ joins vertices $u, v \in V$, i.e., $e \in E$. Graph $G$ can be described through a $|V| \times |V|$ adjacency matrix $\mathbf{A}$ whose generic entry is the edge weight $w_{ \{ u, v \} }$ between $u$ and $v$, where $u, v = 1, \dots, |V|$. In this work, the distance $\delta_4 (\cdot)$ between graph-structured observations $X_{ji} = G_i$ and $X_{jl} = G_l$ is the edge difference distance \citep{hammond2013}, which is defined as the Frobenius norm of the difference between the two adjacency matrices. In symbols,

\begin{equation*}
\delta_4 (G_i, G_l) = || \mathbf{A} ^ i - \mathbf{A} ^ l ||_F = \sqrt{\sum_u \sum_v \Big| e_{ \{ u, v \}} ^ i - e_{ \{ u, v \} }^ l \Big| ^ 2},
\end{equation*}

\noindent where $|| \cdot ||_F$ denotes the Frobenius norm, $\mathbf{A} ^ i$ and $\mathbf{A} ^ l$ are the adjacency matrices of $G_i$ and $G_l$ respectively, and $e_{ \{ u, v \}} ^ i$ and $e_{ \{ u, v \}} ^ l$ are the corresponding generic entries. Edge difference distance ensures wide applicability to binary, signed, weighted, and directed graphs, while maintaining computational efficiency and providing reasonable results in various settings.

\subsection{Coefficient Expansion}

Each type of structured covariate requires a specific coefficient expansion method when performing the splits through feature vector extraction. The types of structured covariates considered for this work's applications are functional and in the form of graphs.

A functional data object $f_i(\cdot)$ can be represented by real-valued components using a basis, which is a set of linearly independent vectors $\bm{\phi}_s(\cdot)$, with $s = 1, \dots, S$. Bases allow approximating arbitrarily well the function as a linear combination 

\begin{equation}\label{basis_expansion}
f_i(t) \approx \sum_{s = 1} ^ S c_{is} \bm{\phi}_s(t),
\end{equation}

\noindent where $c_{is}$ is the coefficient of the $s$-th element of the basis for $f_i(\cdot)$. Since bases are usually such that the vector $\bm{c}_i = (c_{i1}, \dots, c_{iS})$ fully specifies the data object, $\bm{c}_i$ itself can be naturally used as the set of components.

Using the notation of Section \ref{split_featvec_section}, suppose that splitting variable $X_{j^*}^{\mathbf{w}}$ is functional, meaning that its generic element $X_{j^*k}$ is a function $f_k(\cdot)$. The transformation $g_j(\cdot)$ used for any functional covariate is such that 

\begin{equation*}
g_j (f_k) \mapsto \bm{c}_k = (c_{k1}, \dots, c_{k s_j})^T,
\end{equation*}

\noindent where $\bm{c}_k$ is derived by representing $f_k(\cdot)$ using Equation \eqref{basis_expansion} with a basis of $s_j$ elements.

Energy trees use cubic B-splines \citep{ramsay2005} as the basis with generic element $\bm{\phi}_s(\cdot)$ in Equation \eqref{basis_expansion}. Splines are commonly used for approximating non-periodic functional data because they are smooth, efficient, flexible, and parsimonious \citep{ramsay2005}. Cubic splines correspond to the lowest order that guarantees two continuous derivatives, meaning that both the representation of the function and its first derivative are smooth. Among splines, the B-splines basis system is the most popular \citep{ramsay2005}. 

Techniques for performing coefficient expansion on graphs are less established than those for functional data. For the simplest case of undirected (symmetric adjacency matrix) and unweighted (binary adjacency matrix) graphs, energy trees employ the notion of $k$-core decomposition \citep{seidman1983}. The method adequately captures the graph's global connectivity structure \citep{seidman1983,carmi2007}, and is based on transforming the graph into a set of $k$-cores. The $k$-core of a graph $G$, denoted as $C_k(G)$, is defined as the maximal subgraph where every vertex has at least degree $k$. 

To obtain a single vector from the set of $k$-cores, energy trees use the definition of \textit{shell index} \citep{carmi2007}: a vertex $v \in G$ has shell index $i$ if $v \in C_i(G)$, but $v \notin C_{i+1}(G)$. In other words, the shell index of a vertex $v$ represents the highest core to which $v$ belongs. 
Representing $k$-cores as shell indices reduces dimensionality, and these indices can be collected in a single vector that characterizes the entire graph. Specifically, a graph $G$ can be represented using a $|V|$-dimensional vector whose $j$-th entry is the number $s_j$ of vertices of $G$ that have shell index $j$, for $0 \leq j \leq |V|-1$. Such a vector is called \textit{shell distribution} of the graph $G$ and can be denoted as $\bm{s}(G)$.

Suppose that the splitting variable $X_{j^*}^{\mathbf{w}}$ is graph-structured, with the generic $X_{j^*k}$ being a graph $G_k$. The transformation used for any covariate in the form of graphs is such that

\begin{equation*}
g_j(G_k) \mapsto \bm{s}_k = (s_{k1}, \dots, s_{k|V_k|})) ^ T,
\end{equation*}

\noindent where $\bm{s}_k \equiv \bm{s}(G_k)$, and $V_k$ is the set of vertices of graph $G_k$.

The notion of $k$-core was originally introduced for unweighted and undirected graphs but has been extended to both the weighted and the directed cases with $s$-cores \citep{eidsaa2013} and $d$-cores \citep{giatsidis2013}, respectively. On the other hand, the definitions of shell index and shell distribution remain the same. This ensures that energy trees can transform splitting variables in the form of weighted or directed graphs into real-valued components.

\section{Comparison With Traditional Competitors}
\label{app:comparison}

Since energy trees are introduced to overcome the limitations of traditional models in handling structured covariates, comparing their performance in simulation settings is not a straightforward task. One approach is to use feature vector extraction to transform any structured covariate into Euclidean features that competitors can handle. However, this raises two problems. Firstly, the number $s_j$ of components deriving from feature vector extraction is, in general, different for each covariate $j$. Secondly, the number of components for the same covariate may differ across the simulation runs due to specific variable realizations. To address these challenges, it is necessary to correct the relative frequencies of selection of any component accounting for these variations.


The same procedure applies to the unbiasedness analysis, where the root split is forced by removing any stop criterion, and to the power and conditional probability analyses, where splits are not necessarily performed. In the latter case, it suffices to condition the analysis on the actual selection of any covariate.

Let $S$ represent the event of selecting a given component, and $A$ denote the event of that component being available. Since $S \subset A$, the conditional probability of selecting a specific component
is $P(S|A) = P(S) / P(A)$. This implies that estimating $ P(S|A)$ requires dividing the relative frequency of selection of the component by the corresponding relative frequency of availability. Consequently, the relative frequency of selection for the covariate can be obtained by averaging these quantities across all components.

The notation introduced in Section \ref{split_featvec_section} can be used to formalize these ideas. Recall that $\bm{b}_s^j$ represents the $s$-th component resulting from the transformation of the $j$-th variable $X_{j}^{\mathbf{w}}$ via $g_j$. Note that $s = 1, \dots, s_j$, since the number of components depends on the specific transformation $g_j$, and $j = 1, \dots, J$. The estimated probability of selecting the $j$-th covariate $X_{j}^{\mathbf{w}}$ is obtained as a weighted average

\begin{equation}\label{p_tilde}
    \tilde{p}_j = \frac{1}{s_j} \sum_{s = 1}^{s_j} \frac{\hat{f}^j_s}{\hat{a}^j_s},
\end{equation}

\noindent where $\hat{f}^j_s$ is the relative frequency of selecting feature $\bm{b^j_s}$, and $\hat{a}^j_s$ is the relative frequency of availability for feature $\bm{b^j_s}$.

However, summing $\tilde{p}_j$ from Equation \eqref{p_tilde} across $j = 1, \dots, J$ does not necessarily yield $1$. A solution is normalizing each $\tilde{p}_j$ to obtain the proper relative frequencies of selection of the $j$-th covariate,

\begin{equation}\label{p_hat}
    \hat{p}_j = \tilde{p}_j \bigg/ \sum_{j = 1}^J \tilde{p}_j.
\end{equation}

\noindent The quantities $\hat{p}_j$ from Equation \eqref{p_hat}, with $j = 1, \dots, J$, are then used to compare the performance of traditional models with that of energy trees.

\addcontentsline{toc}{chapter}{Bibliography}
\bibliography{bibliography}

\end{document}